\documentclass[twocolumn,showpacs,showkeys,pre,floatfix]{revtex4}
\usepackage{bm}
\usepackage{graphicx}
\usepackage[dvips]{color}
\usepackage{amsmath, amsthm, amssymb}
\usepackage{theorem}
\usepackage{ascmac}
\begin{document}

\definecolor{BrickRed}{rgb}{0.6,0.08,0.4}
\renewcommand{\revised}{\textcolor{BrickRed}}

\title{Design of Oscillator Networks with Enhanced Synchronization Tolerance against Noise}

\author{Tatsuo Yanagita}
\affiliation{Osaka Electro-Communication University, Neyagawa
572-8530, Japan}
\email{yanagita@isc.osakac.ac.jp}
\author{Alexander S. Mikhailov}
\affiliation{Fritz-Haber-Institut der Max-Planck-Gesellschaft, Faradayweg 4-6, 14195
Berlin, Germany}

\begin{abstract}
Can synchronization properties of a network of identical oscillators in the presence of noise be improved through appropriate rewiring of its connections? What are the optimal network architectures for a given total number of connections? We address these questions  by running the optimization process, using the stochastic Markov Chain Monte Carlo method with replica exchange to design the networks of phase oscillators with the increased tolerance against noise. As we find, the synchronization of a network, characterized by the Kuramoto order parameter, can be increased up to 40 \%, as compared to that of the randomly generated networks, when the optimization is applied. Large ensembles of optimized networks are obtained and their statistical properties are investigated. 
\end{abstract}
\date{\today }

\maketitle

\newcommand{\singlefiguresize}{0.85\columnwidth}
\newcommand{\doublefiguresize}{1.85\columnwidth}







\section{Introduction}

Synchronization phenomena are ubiquous in various fields of science and play an important
role in functioning of living systems \cite{Kurths01}.
In the last decade, much interest has been attracted to studies of complex
networks consisting of dynamical elements involved in a
set of interactions \cite{RevModPhys.74.47,Boccaletti2006175}. 
Particular attention has been paid to problems of synchronization in
network-organized oscillator systems \cite{Manrubia04,Arenas08}. 
Investigations focused on understanding the relationship between the
topological structure of a network and its collective synchronous behavior 
\cite{Boccaletti2006175}. Recently, synchronization properties of systems
formed by phase oscillators on static complex networks, such as small-world
networks \cite{Hong02} and scale-free networks \cite{Ichinomiya04,Lee05},
have been considered. It has also been shown that the ability of a network
to give rise to synchronous behavior can be greatly enhanced by exploiting
the topological structure emerging from the growth processes \cite%
{PhysRevE.71.016116,PhysRevLett.94.138701}. However, full understanding of
how the network topology affects synchronization of specific dynamical units
is still an open problem.

One possible approach is to use evolutionary learning mechanisms in order to
construct networks with prescribed dynamical properties. Several models have
been explored, where dynamical parameters were modified in response to the
selection pressure via learning algorithms, in such a way that the system
evolved towards a specified goal \cite{Mikhailov02,Moyano01,yanagita10}. 
This approach can also be employed to design phase oscillator networks with desired 
synchronization properties.  
Using heterogeneous oscillators with a dispersion of natural frequencies, we have previously shown how these elements can be optimally connected, by using a given number of links, so that the best synchronization level is achieved \cite{yanagita10}.

Here, our attention is focused on synchronization enhancement in networks of identical phase oscillators in the presence of noise. 
In such systems, noise acting on the oscillators competes with the coupling which favors the emergence of coherent dynamics \cite{Kuramoto84,Manrubia04}. 
The question is how to connect a set of phase oscillators, so that the resulting network  exhibits the strongest possible synchronization despite the presence of noise, under the constraint that the total number of available links and, thus, the mean connectivity are fixed.

To design optimal networks, stochastic Markov Chain Monte Carlo (MCMC) method with replica exchange  \cite{yanagita10} is
used by us. Large ensembles of optimal networks are constructed and their
common statistical properties are analyzed. 
As we observe, the typical
structure of a synchronization-optimized network is strongly dependent on
its connectivity. Sparse optimal networks, with a small number of
links, tend to display a star-like structure. 
As the connectivity is increased, synchronization-optimized networks show a
transition to the architectures with interlaced cores.

The paper is organized as follows. In Sec.~\ref{sec:model}, we introduce a
model of identical phase oscillators occupying nodes of a directionally
coupled network and define the synchronization measure for this system. The
optimization method is also introduced in this section. 
Construction of optimized networks and their statistical analysis are performed in Sec.~\ref{sec:numerical}. The results are finally discussed in Sec.~\ref{sec:summary}


\section{The Model and the Optimization Method}
\label{sec:model}

For identical oscillators, it is known that, in absence of noise, even very weak coupling can lead to complete synchronization \cite{Kuramoto84,RevModPhys77137}.  
Below, we consider the effects of noise acting on a network of coupled identical phase oscillators, so that the model equations are
\begin{equation}
\frac{d \theta_i}{dt}=\omega_0+\frac{\lambda}{N} \sum_{j=1}^{N}w_{j,i}\sin(\theta_j-\theta_i)+\xi_i(t),
\label{eq:model}
\end{equation}
where $\xi_i(t)$ are independent white noises, such that $\langle \xi_i(t) \rangle=0$ and $\langle \xi_i(t) \xi_j(t')\rangle=S^2 \delta_{i,j} \delta(t-t')$.
Interactions between the oscillators are specified by the matrix $\mathbf{w}$ with the elements $w_{i,j} = 1$, if there is a connection, and $w_{i,j} = 0$ otherwise. Generally, the connection matrix is asymmetric.
Note that since the rotation frequencies of all oscillators are the same, we can always go into the rotational frame $\theta_i \mapsto \theta_i-\omega_0 t$ and thus eliminate the term with $\omega_0$.
Hence, without any loss of generality one can set $\omega_0=0$ in Eqs.~(\ref{eq:model}).
It is known that, for global coupling, this model shows a transition to synchronization as the ratio of the coupling strength to the noise intensity is increased (see, e.g., \cite{Mikhailov06}).

To quantify synchronization of the oscillators, global phase  
\begin{equation}
r(t)=\frac{1}{N} \sum_{i=1}^{N}\exp (\mathbf{i}\theta _{i})
\end{equation}%
will be employed. 
To measure the degree of synchronization, we numerically integrate Eq.~(\ref{eq:model}) with the initial conditions $\theta _{i}(t=0)=0$ and calculate the average of $|r(t)|$ over a long time $T,$ 
\begin{equation}
R(\mathbf{w})= \frac{1}{T}\int_{0}^{T}\left\vert r(t)\right\vert dt.
\end{equation}%

Our aim is to determine the network $\mathbf{w}=\{w_{i,j}\}$ which would exhibit the
highest degree of synchronization, provided that the total number $K$ of links
is fixed and the noise intensity $S$ is given. The network construction
can be seen as an optimization problem. The optimization task is to maximize
the order parameter and, possibly, bring it to unity by changing the network 
$\mathbf{w}$. 


To study statistical ensembles of optimized networks, the MCMC method \cite{Landau05,Newman99,Liu01}, which has previously been applied to dynamical systems \cite{Cho94,Bolhuis98,Vlugt00,Kawasaki05,Sasa06,Giardin06,Tailleur07,yanagita09,yanagita10},
will be used by us. 

We sample networks from the ensemble with the Gibbs distribution $P(\mathbf{w}) \sim \exp (\beta R(\mathbf{w}))$ by the MCMC method.
To improve the sampling efficiency, we use the Replica Exchange Monte Carlo (REMC) algorithm, and the details of the algorithm can be seen in ~\cite{yanagita10}.

We mainly consider the canonical ensemble average of a network function $f(\cdot )$,i.e.,  
\begin{equation}
\langle f  \rangle_{\beta}=\sum_{w}\frac{f(\mathbf{w})\exp (\beta R(\mathbf{w}))}{Z(\beta )},  \label{eq:gibbs}
\end{equation}
where $Z(\beta )=\sum_{w}\exp (\beta R(\mathbf{w}))$ is the partition function and the parameter $\beta $ plays the role of the inverse temperature. 

\section{Numerical Investigations}
\label{sec:numerical}

To determine the synchronization degree of a given network at each iteration
step of the optimization procedure, equations~(\ref{eq:model}) were
numerically integrated with the time increment $\Delta t=0.01.$ 
Due to limited computational resources, only relatively small oscillator ensembles of sizes $N=15$ are considered in this study. 
The noise intensity is always $S=0.3$.

Initial phases are $\theta _{i}(0)=0$. 
Hence, the order parameter at $t=0$ is always equal to unity. 
To construct an initial random network with a given number $K$ of
connections and, thus, with given connectivity $p=K/N(N-1)$, $K$ off-diagonal
elements of the matrix $\mathbf{w}$ are randomly and independently selected
and set equal to unity.

For time averaging, relatively long intervals $T=10000$ were typically
used, since the convergence of the order parameter is slow. 
The results did not significantly depend on $T$ when sufficiently
large lengths $T$ were taken. 

In parallel, evolution of $M+1$ replicas with different inverse temperatures $\beta _{m}=\delta \beta \times m, m=0,1, \dots ,M$ has been performed ($M=63$ and $\delta \beta =5$). 
The statistical results did not significantly depend on the particular choice of inverse temperatures.

At every five Monte Carlo steps (mcs), the performances of a randomly chosen pair of
replicas were compared and exchanged, as described above. For display and
statistical analysis, sampling at each every 50 mcs after a transient of $%
5000$ mcs has been undertaken.

\subsection{Optimization at different temperatures}

Synchronization-optimized networks were obtained by running  evolutionary
optimization. In this process, the order parameter was progressively
increasing until a saturation state has been reached. Figure~\ref%
{fig:timeseries} gives examples of the optimization processes at different
temperatures. As clearly seen, when using replicas with the larger inverse
temperature $\beta$,  larger values of the order parameter could be
reached, although the optimization process was then slower. 
This suggests that, for the considered problem, the replicas do not actually get trapped in the local minima even at large $\beta$ and that already such low-temperature replicas can be efficiently used to sample the optimized networks.

\begin{figure}[tbp]
\begin{center}
\resizebox{\singlefiguresize}{!}{\includegraphics{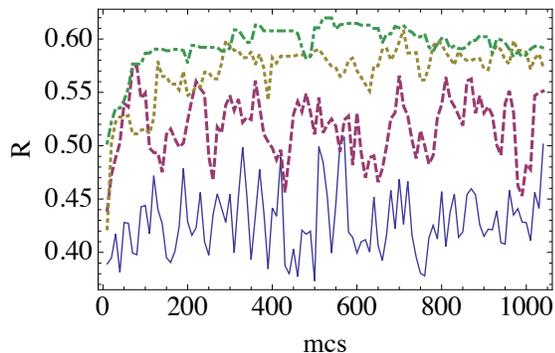} }
\end{center}
\caption{(Color online). Examples of evolution of the synchronization order parameter during the optimization process. The blue solid, red broken, yellow dotted, green dotted dash curves are for the inverse temperatures $\protect\beta =\protect\beta _{0},\protect\beta _{16},\protect\beta _{32}$
and $\protect\beta _{M}$, respectively. 
The blue solid line ($\protect\beta _{0}=0$) corresponds to the networks
generated by only random rewiring. The parameters are $p=0.1,\protect%
\lambda =1.0,\protect\gamma =0.3,M=63,\protect\delta \protect\beta =5$.}
\label{fig:timeseries}
\end{figure}

\begin{figure*}[tbp]
\begin{center}
\resizebox{\doublefiguresize}{!}{\includegraphics{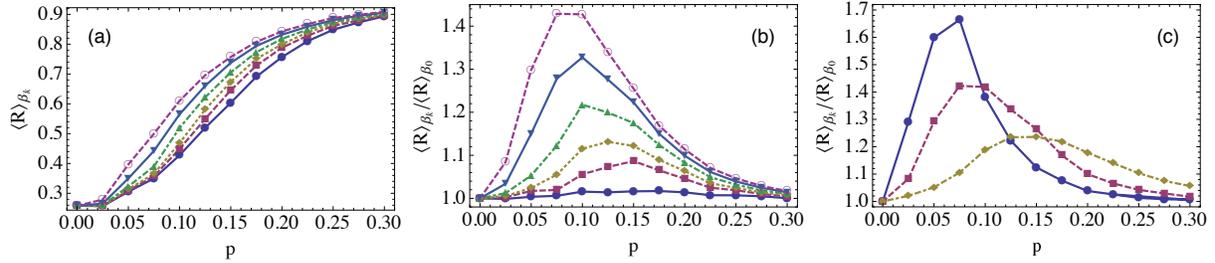} }
\end{center}
\caption{(Color online). Average synchronization order parameters (a) and ratios of the order parameters (b) as functions of the network connectivity $p$. The blue filled circles are for the replica  $\protect%
\beta _{0}$, i.e., the ensemble of  randomly rewired networks. The red squares,
yellow diamonds, green triangles, blue inverted triangles, and red open circles are for the replicas
with $\protect\beta =\beta_{0},\beta_{4},\beta_{8},\beta_{16}, \beta_{32}$ and $\beta_{M}$, respectively. 
(c) Noise intensity dependence of the ratios of the order parameters for $\beta_{M}$ is shown for $S=0.2$ (blue filled circles), $S=0.3$ (red squares), and $S=0.4$ (yellow diamonds).
Other parameters are the same as in Fig.~\protect\ref{fig:timeseries}}
\label{fig:order}
\end{figure*}

After the transients, statistical averaging of the order parameter over the ensemble
with the Gibbs distribution has been performed, according to Eq.~(\ref{eq:gibbs}).
In Fig. ~\ref{fig:order}(a), the averaged order parameter $\langle R \rangle_{\beta}$ is
displayed as a function of the connectivity $p$ for several different
inverse temperatures $\beta $. 
The blue solid circle symbols  show the averaged
order parameter corresponding to the replica with $\beta _{0}=0,$ i.e. for
an infinitely high temperature. 
We see that the averaged order parameter
increases with the network connectivity $p$ even if the networks are
produced by only random rewiring. The red open circles show the average order
parameters for the ensemble corresponding to the replicas with the lowest
inverse temperature $\beta _{M}$. 
Generally, greater order parameters can be obtained by running evolution at higher inverse temperatures  $\beta $. 
At each connectivity $p$, the order parameter
is gradually increased with increasing $\beta $ and is approximately
saturated at $\beta _{M}$. This means that, even if one further increases $%
\beta $, only slight improvements of the averaged order parameter can be
expected. Thus, the networks sampled by the replica with the largest inverse
temperature $\beta _{M}$ are already yielding a representative optimal
ensemble.

Figure~\ref{fig:order}(b) shows the ratio $\langle R \rangle_{\beta _{M}}/\langle R \rangle_{\beta _{0}}$ of the order parameters averaged over network ensemble with the highest inverse temperature $\beta _{M}$ and with the zero inverse temperature (i.e. the ensemble with purely random rewiring) for different connectivities p.
Since there is no room for the improvement of the order parameter when the number of links is small, the ratio tends to unity as the connectivity  $p$ is decreased. On the other hand,  when $p=1$, global coupling is realized, for which, under the chosen coupling strength, full synchronization occurs.
As evidenced by this Figure, the difference between the synchronization
capacities of the optimized and random networks is most pronounced at the
intermediate connectivities, for $p$ around 0.1.
The noise intensity dependence for the synchronization capacities is shown in Fig~\ref{fig:order}(c).
When the noise intensity is small, the ratio becomes larger and the maximum is shifted to the smaller connectivities $p$.

\subsection{Collective dynamics}

\begin{figure}[tbp]
\begin{center}
\resizebox{\singlefiguresize}{!}{\includegraphics{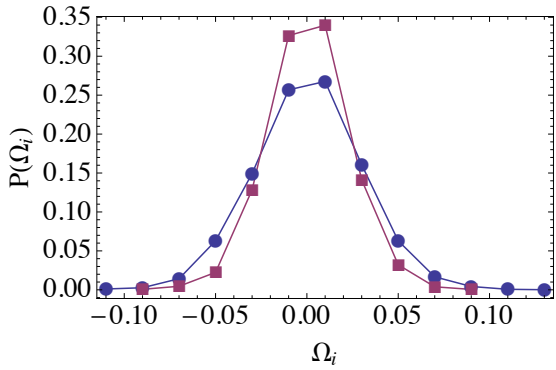} }
\end{center}
\vspace{0mm}
\caption{(Color online). Distributions of the winding number for the ensembles of 500 realizations of random rewiring networks (sampled by replica with $\beta_0$) and of synchronization-optimized networks  (sampled by replica with $\beta_M$).
The blue circles are for random networks and the red squares are for the synchronization-optimized ones. 
The parameters are same as in Fig.~\protect\ref{fig:timeseries}.
}
\label{fig:winding}
\end{figure}

To analyze differences in the collective dynamics of phases oscillators in random and  synchronization-optimized networks, we have calculated the winding number of each oscillator, $\Omega_i=\frac{1}{T}(\theta_i(T)-\theta_i(0))$ for many realizations of random (sampled by replica with $\beta_0$) and synchronization-optimized  (sampled by replica with $\beta_M$) networks, and determined the probability distributions of winding numbers for both ensembles. 
As shown in Fig.~\ref{fig:winding}, there is a significant difference between these two distributions .
The probability peak at $\Omega_i=0$ for the synchronization-optimized ensemble is higher and more narrow than that for the random-rewiring ensemble.
This means that synchronization-optimized networks tend to have more elements oscillating with the common frequency in the presence of the external noises, as compared with random rewired networks.
Thus, elements in the synchronization-optimized network behave more coherently than those in a random network.

\subsection{Architectures of synchronization-optimized networks}

\begin{figure*}[tbp]
\begin{center}
\resizebox{\doublefiguresize}{!}{\includegraphics{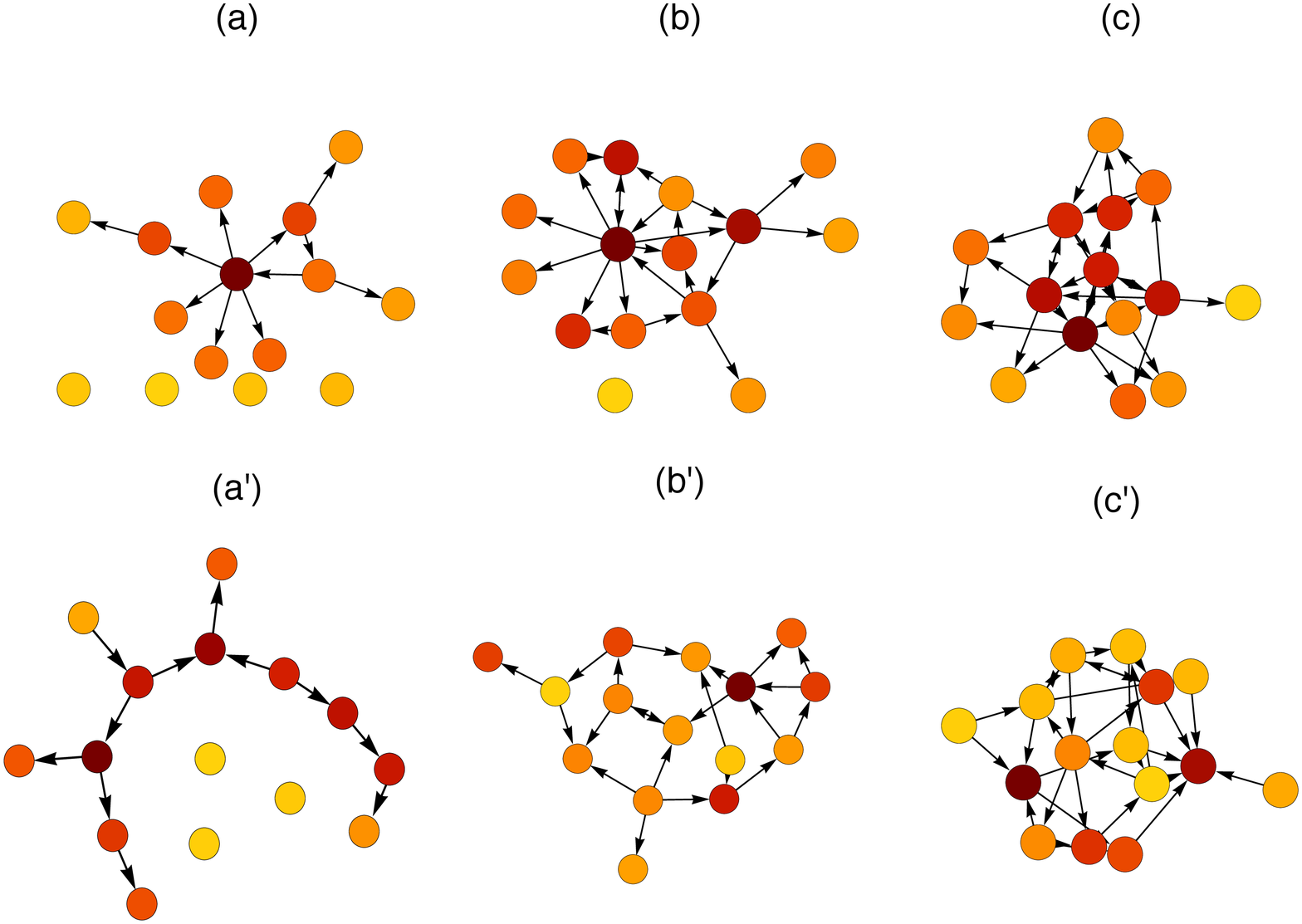} }
\end{center}
\vspace{-5mm}
\caption{(Color online). Examples of synchronization-optimized networks with different connectivities (a) $p=0.05$, (b) $p=0.1$, and (c) $p=0.15$.  (a'), (b') and (c') correspond to typical random networks with the same connectivity for comparison. 
The nodes are colored according to the phase correlation $\eta_i$, and the darker color indicates  an oscillator having stronger correlations with the global order. 
The other parameters are same as in Fig.~\protect\ref{fig:timeseries}.
}
\label{fig:typical_graph}
\end{figure*}

Several typical synchronization-optimized networks are shown in Fig.~\ref{fig:typical_graph}. 
Their structures strongly depend on the number of available connections (the number of links is always conserved during an optimization process). 
When connectivity $p$ is small [Fig.~\ref{fig:typical_graph} (a)], designed networks usually have star structures.
The central element acts on a group of periphery elements which have no connections among them. 
Additionally, a number of disconnected elements are present. 
If a larger number of links is available  [Fig.~\ref{fig:typical_graph} (b)], a core, formed by a group of interconnected elements, becomes formed. 
There are also periphery elements, which are affected by the core, but do not influence its dynamics.
 As the mean connectivity of the network is increased, the core grows at the expense of the periphery elements. 
 Thus, the network starts to include [Fig.~\ref{fig:typical_graph} (c)] a relatively large group of highly connected elements, with only a few elements which are loosely connected and belong to the periphery.

For a synchronization-optimized network, we have integrated the equations~(\ref{eq:model})
 for a long time, and calculated the correlations $\eta_i$ between the phase of a local oscillator $\theta_i$ and that of the global order variable $r(t)$ defined as
\[
\eta_i=\frac{1}{T} \left| \int_{0}^{T} r(t) \exp(\mathbf{i} \theta_i)dt \right|.
\]
These quantities show how strongly the dynamics of an oscillator $i$ is synchronized with the global signal $r(t)$. 
The nodes in Fig.~\ref{fig:typical_graph} are colored according to the rescaled values $\eta_i$, i.e., $\{\min_{i} \eta_i+\eta_i\}/\{\max_i \eta_i-\min_{i} \eta_i\}$. 
The darker color indicates an oscillator having the stronger phase correlation with the global signal.

Figure~\ref{fig:typical_graph} suggests that the phases of central oscillators are strongly correlated with the phase of the global signal. 
In order to check this more clearly, we have divided all oscillators into the groups with equal degrees and separately determined average correlations with the global signal for each group. Thus, quantities $\eta_k$ have been calculated,
\[
\eta_k=\frac{1}{\sum_{i=1}^{N} \delta_{k,k_i}} \sum_{i=1}^{N} \eta_i \delta_{k,k_i},
\]
where $k_i$ denotes the total degree of a node $i$, i.e., 
\[
k_i=k^{+}+k^{-}, \;
k_{i}^{+}=\sum_{j=1}^{N} w_{i,j}, \;
k_{i}^{-}=\sum_{j=1}^{N} w_{j,i}, 
\]
with $k^{+}$ and $k^{-}$ being the ingoing and outgoing degrees, respectively.

\begin{figure}[tbp]
\begin{center}
\resizebox{\singlefiguresize}{!}{\includegraphics{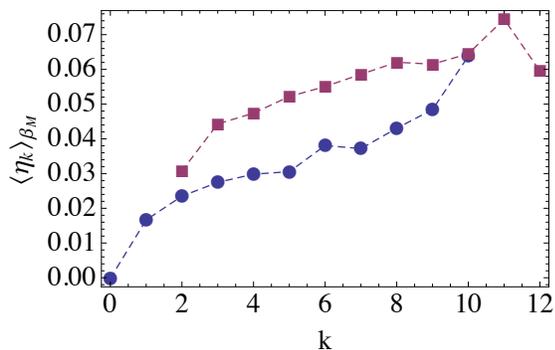} }
\end{center}
\vspace{0mm}
\caption{(Color online). Correlation between phases of the global order and local oscillators as a function of the degree. 
Averaging over 500 realizations sampled by replica with $\beta_M$.
The blue circles and red squares are for $p=0.05$ and $p=0.20$, respectively. 
The other parameters are same as in Fig.~\protect\ref{fig:timeseries}.
}
\label{fig:zcorr}
\end{figure}

In Figure~\ref{fig:zcorr}, phase correlations $\eta_k$, averaged over an ensemble of synchronization-optimized networks, are plotted as a function of the degree $k$ for different network connectivities $p$. 
We see that, on the average, nodes with higher degrees are stronger correlated with the global signal.
Thus, the oscillators having many connections act as organizing centers of the synchronization. 
Furthermore, as seen in Fig.~\ref{fig:zcorr}, phase correlations for the nodes with the same degree become larger as the connectivity is increased.
This tendency can be understood if we take into account that the synchronization-optimized networks usually have shallow tree-like structures for the smaller connectivities $p$.
As $p$ increases,  the network becomes interlaced and has many loops [Fig.~\ref{fig:typical_graph}(c)]. 
Since the feedback in a loop enhances the correlation, the averaged phase correlation of nodes with the same degree becomes larger as $p$ increases.

Note that in a star structure, the central node does not receive any signal from other oscillators; thus, the phase of the oscillator in the center is only affected by the applied noise.
On the other hand, when outgoing connections from the center to the periphery elements are present, the central oscillator effectively acts as a source of common noise applied to the peripheral nodes.
Recently, it has been shown that  common noise can induce  synchronization in an ensemble of identical oscillators \cite{Teramae04,Nakao07}. 
This phenomenon may be responsible for the development of correlations between the peripheral elements and the central oscillator. Similar behavior may take place when, instead of a single central node, a core of highly connected oscillators is present in a network.

\subsection{Degree distributions}

\begin{figure*}[tbp]
\begin{center}
\resizebox{\doublefiguresize}{!}{\includegraphics{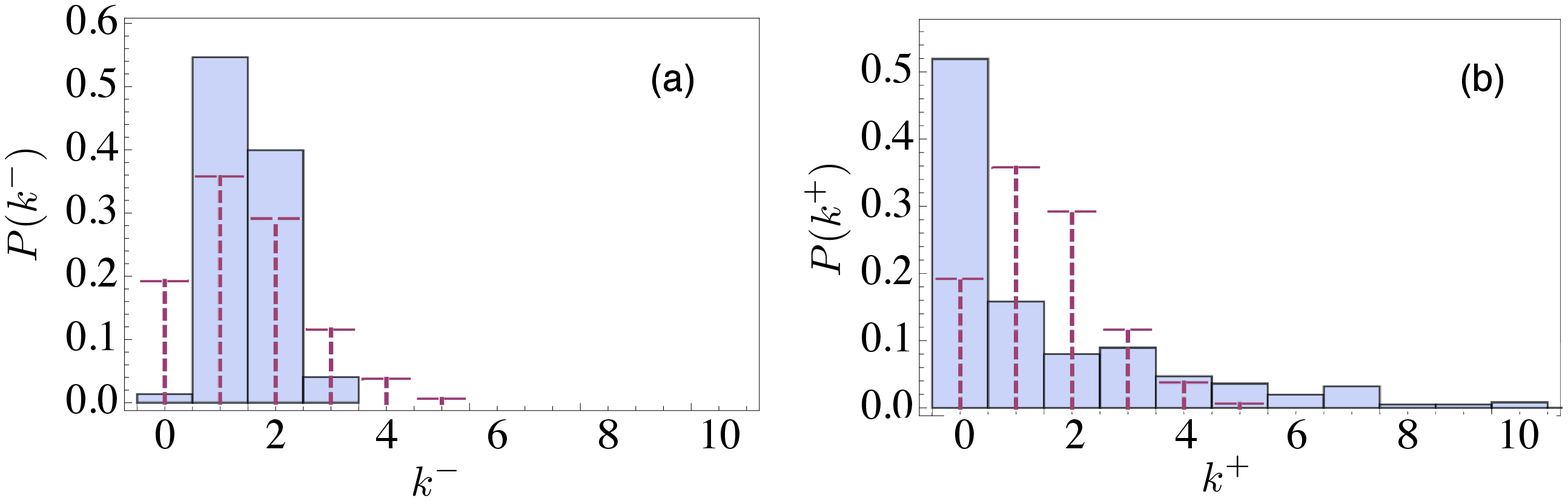} }
\end{center}
\vspace{0mm}
\caption{(Color online). The distributions of ingoing (a) and outgoing (b) degrees for random and synchronization-optimized networks. Each distribution is averaged over 200 network realizations. 
The blue bars show the distribution for synchronization-optimized networks, whereas the red dashed lines are for the random networks. 
The parameters are same as in Fig.~\protect\ref{fig:timeseries}.
}
\label{fig:degree}
\end{figure*}

To statistically investigate architectures of designed networks, ingoing and outgoing degrees of their nodes have been considered.
By sampling over $200$ realizations from synchronization-optimized ensemble, we have obtained the ingoing and outgoing degree distributions at $p=0.10$, as shown in Fig.~\ref{fig:degree}.
For the ensemble of random rewiring networks,  both ingoing and outgoing degrees obey the same Poisson distribution (red broken lines in the figure represent the in- and out-degree distributions of networks sampled by the replica with $\beta_0$). 
As clearly seen in Fig.~\ref{fig:degree}, most of nodes in the synchronization-optimized networks have only one ingoing connection and no outgoing connections.
This indicates that many periphery nodes exist, consistent with a typical realization of synchronization-optimized network shown in Fig.~\ref{fig:typical_graph}(a). 
%
Moreover, the outgoing degrees of synchronization-optimized networks are distributed more broadly than those of random rewiring networks, i.e., a long tail in the outgoing connection distribution has emerged.
This reflects the development of core nodes.
Hence, there are two principal types of nodes, i.e., core and periphery nodes, in the synchronization-optimized networks.
The core nodes have many outgoing connections and a smaller number of ingoing connections, whereas the periphery nodes tend to have small numbers of ingoing connections.

\begin{figure}[tbp]
\begin{center}
\resizebox{\singlefiguresize}{!}{\includegraphics{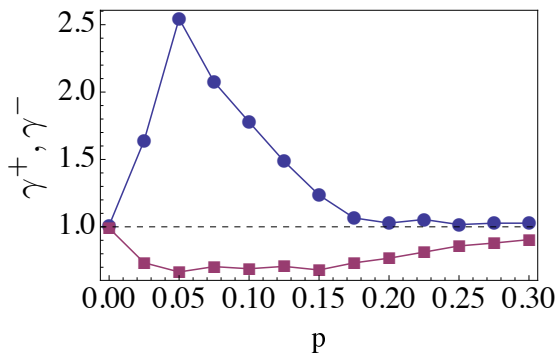} }
\end{center}
\vspace{0mm}
\caption{(Color online). 
The dependences of relative maximum in- and out-going degrees of synchronization-optimized networks on their mean connectivity $p$. 
The data for out- and in-degrees are shown by blue circles and red squares, respectively. 
Averaging over 500 realizations of synchronization-optimized and random networks.
The parameters are the same as in Fig.~\protect\ref{fig:timeseries}.
}
\label{fig:maxdegree}
\end{figure}

In order to further investigate the statistics of network structures as a function of the network connectivity, we have calculated the maximum of ingoing and outgoing degrees of each synchronization-optimized network, $k^{+}_{\max}=\max_{i} (k_i^{+})$ and $k^{-}_{\max} =\max_{i} (k_i^{-})$, respectively, and averaged them over many realizations. 
In Fig.~\ref{fig:maxdegree}, the ratios of averaged maximum ingoing and outgoing degrees of the synchronization-optimized networks to those of the random networks, i.e., $\gamma^{+}=\langle k^{+}_{\max} \rangle_{\beta_M}/\langle k^{+}_{\max} \rangle_{\beta_0}$ and $\gamma^{-}=\langle k^{-}_{\max} \rangle_{\beta_M}/\langle k^{-}_{\max} \rangle_{\beta_0}$, are shown.
As $p$ increases, the ratio of the averaged maximum outgoing degree of synchronization-optimized networks to that of the random-rewiring networks increases steeply and takes the maximum in the vicinity of $p_c = 0.075$,  while that of the outgoing degree (shown by red square symbols) decreases and takes the minimum at approximately the same $p_c$. 
In the vicinity of $p_c$, the nodes with a small number of ingoing connections and a large number of outgoing connections (corresponding to the cores) are found in the synchronization-optimized networks.

\subsection{Eigenvalues of the Laplacian matrix}
\renewcommand\Re{\operatorname{Re}}
\renewcommand\Im{\operatorname{Im}}

\begin{figure}[tbp]
\begin{center}
\vspace{5mm}
\resizebox{\singlefiguresize}{!}{\includegraphics{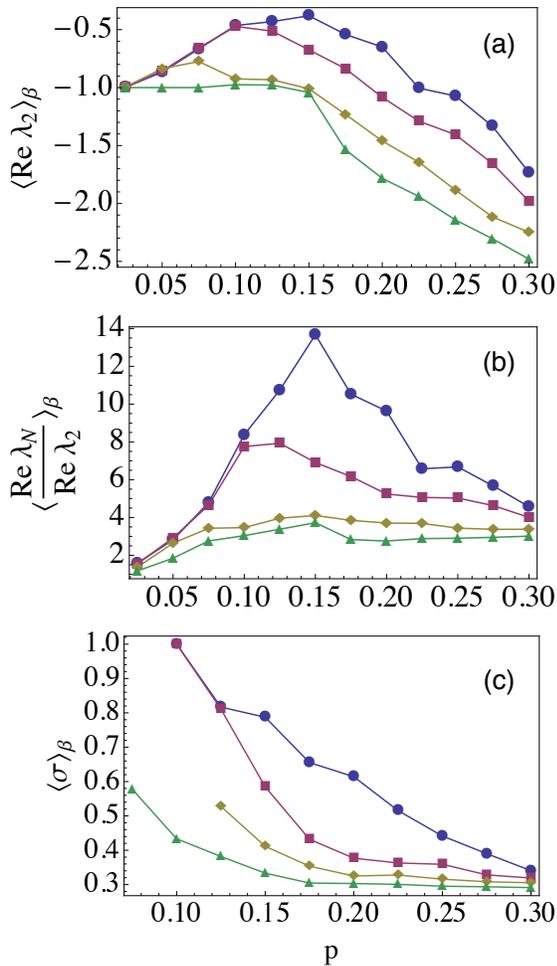} }
\end{center}
\vspace{0mm}
\caption{(Color online). (a) $\Re \lambda_2$,  (b) $\Re \lambda_n/\Re \lambda_2$, and (c) $\sigma$ averaged over the Gibbs ensemble at different inverse temperatures $\beta$.
The circles indicate average over random networks. 
The squares are averages over the Gibbs ensemble with $\beta=\beta_8$, the diamonds are with $\beta_{32}$ and triangles are averages with, i.e., $\beta=\beta_M$. 
The other parameters are same as in Fig.~\protect\ref{fig:timeseries}.
}
\label{fig:eigen}
\end{figure}

 The Laplacian matrix $\mathbf{L}$ for network $\mathbf{w}$ is defined as
\begin{equation}
L_{i,j}=
  \begin{cases}
w_{i,j} &(i\neq j) \\
\displaystyle {-\sum^{N}_{j=1}w_{i,j}} & (i=j).
  \end{cases}
\end{equation}
Since the considered networks are directed, the eigenvalues of their Laplacian matrices are complex. 
We can order the eigenvalues according to the magnitudes of their real parts, i.e. as
\[
0=\Re(\lambda_1)>\Re(\lambda_2)>\cdots>\Re(\lambda_{N}).
\]
The eigenvalues of the Laplacian matrix are known to play an important role for the synchronizability of oscillator networks \cite{Arenas08}.
Therefore, we have computed $\Re \lambda_2$ and $\Re \lambda_N/ \Re \lambda_2$ for many realizations of synchronization-optimized networks.
In Fig.~\ref{fig:eigen} (a), $\langle \Re \lambda_2 \rangle_\beta$ as function of the connectivity is shown for different inverse temperatures $\beta$.
It is clearly seen that $\langle \Re \lambda_2 \rangle_\beta$ decreases with $\beta$.
Since $\Re \lambda_2$ determine the inverse relaxation time to the synchronized state in oscillator networks \cite{Arenas08}, this indicates that the time needed to achieve the synchronized state decreases with $\beta$.
The  ratio $\langle \Re \lambda_N/\Re \lambda_2 \rangle_\beta$ averaged over the Gibbs ensemble with $\beta$ is shown in Fig.~\ref{fig:eigen}(b).
The displayed dependencies reveal that the ratio, which specifies the to synchronizability, decreases as the optimization level, i.e., $\beta$ is increased.

Recently, fluctuations in the collective signal and oscillation precision were also linked to the eigenvalues of
 the Laplacian matrix \cite{Kori201261,Masuda10}.
 The mean intensity of the fluctuations of the collective signal in an ensemble of components subject to independent Gaussian noises can be estimated by the norm of left eigenvector $v$, corresponding to the zero eigenvalue, $v L$=0 and normalized as $\sum_{i=1}^{N} v_i=1$ (see details in \cite{Masuda10}).
When all independent Gaussian noises have the same strength, the mean-square dispersion of the collective signal can be estimated as $\sigma=\sqrt{\sum_{i=1}^{N} v_i^2}$.
We have computed this property for the ensembles of our designed networks.
In Fig.~\ref{fig:eigen}(c), we have shown $\sigma$ as a function of the network connectivity for different optimization levels, i.e., different $\beta$. 
As we see, $\sigma$ decreases with $\beta$, implying that collective fluctuations are suppressed through the optimization.
We have also estimated the oscillation precision \cite{Kori201261} for our synchronized-optimized networks. 
The results indicate a tendency to increase the precision at higher optimization levels (the figure does not shown in the figure).

While our networks have been only optimized with respect to their synchronization ability in the presence of noise, the above analysis clearly shows that the designed networks turn out to be also optimized with respect to a number of other properties. For the designed networks, the time of relaxation to the synchronized state in the absence of noise is shorter. In the presence of weak noise, such networks have lower intensity of fluctuations in the collective signal and higher oscillation precision.

\section{Conclusions}
~\label{sec:summary}
We have designed synchronization-optimized networks with a fixed number of
links for a population of identical oscillators under action of independent external noises. 
This has been done by using the Markov Chain Stochastic Monte Carlo method complemented by the Replica Exchange algorithm.
Large ensembles of networks with improved synchronization properties have been constructed at different mean connectivities and their statistical properties have been analyzed by using various characterization tools.

Our analysis reveals that the architectures leading to the improved synchronization of identical oscillators in the presence of noise are essentially different from the optimal synchronization architectures for heterogeneous oscillator populations without noise, which have previously been studied \cite{yanagita10}.
When the number of available links is small, synchronization-optimized networks are typically star-shaped structures. 
As the number of links grows, the designed networks are seen to develop dense cores, which replace a single central element in the star networks. 
The core expands as the number of available links is increased, and eventually the network becomes strongly interlaced.
%
The star and core-periphery structures of the designed networks can be qualitatively understood, if one takes into account that the central elements in such networks are effectively operating as the source of common noise for the periphery elements. 
It is known  \cite{Teramae04,Nakao07} that common noise can induce synchronization in the populations of disconnected oscillators or, in our case, in the group of periphery elements all connected to the same central elements or a central core.

Thus, we have shown that  efficient design of oscillator networks with the improved synchronization properties is possible. 
The architectures of such optimal networks strongly depend on the constraints, such as the total number of links available. Through the appropriate rewiring of a network, a strong gain in the synchronization signal can be achieved. 
Although our study has been performed for a simple system of phase oscillators, similar evolutionary optimization methods can be applied to construct networks of different origins, where the dynamics of individual oscillators may be significantly more complex. 

\acknowledgments

This study has been partially supported by the Ministry of Education,
Science, Sports and Culture, Grant-in-Aid for Scientific Research (Grant No. 21540376, 22120501)
and the Volkswagen Foundation (Germany).


\bibliography{refkuramoto,refmcmc}

\begin{thebibliography}{31}
\expandafter\ifx\csname natexlab\endcsname\relax\def\natexlab#1{#1}\fi
\expandafter\ifx\csname bibnamefont\endcsname\relax
  \def\bibnamefont#1{#1}\fi
\expandafter\ifx\csname bibfnamefont\endcsname\relax
  \def\bibfnamefont#1{#1}\fi
\expandafter\ifx\csname citenamefont\endcsname\relax
  \def\citenamefont#1{#1}\fi
\expandafter\ifx\csname url\endcsname\relax
  \def\url#1{\texttt{#1}}\fi
\expandafter\ifx\csname urlprefix\endcsname\relax\def\urlprefix{URL }\fi
\providecommand{\bibinfo}[2]{#2}
\providecommand{\eprint}[2][]{\url{#2}}

\bibitem[{\citenamefont{Kurths et~al.}(2001)\citenamefont{Kurths, Pikovsky, and
  Rosenblum}}]{Kurths01}
\bibinfo{author}{\bibfnamefont{J.}~\bibnamefont{Kurths}},
  \bibinfo{author}{\bibfnamefont{A.}~\bibnamefont{Pikovsky}}, \bibnamefont{and}
  \bibinfo{author}{\bibfnamefont{M.}~\bibnamefont{Rosenblum}},
  \emph{\bibinfo{title}{Synchronization: A Universal Concept in Nonlinear
  Sciences}} (\bibinfo{publisher}{Cambridge Univ. Press, Cambridge},
  \bibinfo{year}{2001}).

\bibitem[{\citenamefont{Albert and Barab\'asi}(2002)}]{RevModPhys.74.47}
\bibinfo{author}{\bibfnamefont{R.}~\bibnamefont{Albert}} \bibnamefont{and}
  \bibinfo{author}{\bibfnamefont{A.-L.} \bibnamefont{Barab\'asi}},
  \bibinfo{journal}{Rev. Mod. Phys.} \textbf{\bibinfo{volume}{74}},
  \bibinfo{pages}{47} (\bibinfo{year}{2002}).

\bibitem[{\citenamefont{Boccaletti et~al.}(2006)\citenamefont{Boccaletti,
  Latora, Moreno, Chavez, and Hwang}}]{Boccaletti2006175}
\bibinfo{author}{\bibfnamefont{S.}~\bibnamefont{Boccaletti}},
  \bibinfo{author}{\bibfnamefont{V.}~\bibnamefont{Latora}},
  \bibinfo{author}{\bibfnamefont{Y.}~\bibnamefont{Moreno}},
  \bibinfo{author}{\bibfnamefont{M.}~\bibnamefont{Chavez}}, \bibnamefont{and}
  \bibinfo{author}{\bibfnamefont{D.-U.} \bibnamefont{Hwang}},
  \bibinfo{journal}{Phys. Rep.} \textbf{\bibinfo{volume}{424}},
  \bibinfo{pages}{175 } (\bibinfo{year}{2006}).

\bibitem[{\citenamefont{Manrubia et~al.}(2004)\citenamefont{Manrubia,
  Mikhailov, and Zanette}}]{Manrubia04}
\bibinfo{author}{\bibfnamefont{S.}~\bibnamefont{Manrubia}},
  \bibinfo{author}{\bibfnamefont{A.}~\bibnamefont{Mikhailov}},
  \bibnamefont{and} \bibinfo{author}{\bibfnamefont{D.}~\bibnamefont{Zanette}},
  \emph{\bibinfo{title}{Emergence of Dynamical Order: Synchronization Phenomena
  in Complex Systems}} (\bibinfo{publisher}{World Scientific, Singapore},
  \bibinfo{year}{2004}).

\bibitem[{\citenamefont{Arenas et~al.}(2008)\citenamefont{Arenas, Diaz-Guilera,
  Kurths, Moreno, and Zhou}}]{Arenas08}
\bibinfo{author}{\bibfnamefont{A.}~\bibnamefont{Arenas}},
  \bibinfo{author}{\bibfnamefont{A.}~\bibnamefont{Diaz-Guilera}},
  \bibinfo{author}{\bibfnamefont{J.}~\bibnamefont{Kurths}},
  \bibinfo{author}{\bibfnamefont{Y.}~\bibnamefont{Moreno}}, \bibnamefont{and}
  \bibinfo{author}{\bibfnamefont{C.}~\bibnamefont{Zhou}},
  \bibinfo{journal}{Phys. Rep.} \textbf{\bibinfo{volume}{469}},
  \bibinfo{pages}{93} (\bibinfo{year}{2008}).

\bibitem[{\citenamefont{Hong et~al.}(2002)\citenamefont{Hong, Choi, and
  Kim}}]{Hong02}
\bibinfo{author}{\bibfnamefont{H.}~\bibnamefont{Hong}},
  \bibinfo{author}{\bibfnamefont{M.~Y.} \bibnamefont{Choi}}, \bibnamefont{and}
  \bibinfo{author}{\bibfnamefont{B.~J.} \bibnamefont{Kim}},
  \bibinfo{journal}{Phys. Rev. E} \textbf{\bibinfo{volume}{65}},
  \bibinfo{pages}{026139} (\bibinfo{year}{2002}).

\bibitem[{\citenamefont{Ichinomiya}(2004)}]{Ichinomiya04}
\bibinfo{author}{\bibfnamefont{T.}~\bibnamefont{Ichinomiya}},
  \bibinfo{journal}{Phys. Rev. E} \textbf{\bibinfo{volume}{70}},
  \bibinfo{pages}{026116} (\bibinfo{year}{2004}).

\bibitem[{\citenamefont{Lee}(2005)}]{Lee05}
\bibinfo{author}{\bibfnamefont{D.-S.} \bibnamefont{Lee}},
  \bibinfo{journal}{Phys. Rev. E} \textbf{\bibinfo{volume}{72}},
  \bibinfo{pages}{026208} (\bibinfo{year}{2005}).

\bibitem[{\citenamefont{Motter et~al.}(2005)\citenamefont{Motter, Zhou, and
  Kurths}}]{PhysRevE.71.016116}
\bibinfo{author}{\bibfnamefont{A.~E.} \bibnamefont{Motter}},
  \bibinfo{author}{\bibfnamefont{C.}~\bibnamefont{Zhou}}, \bibnamefont{and}
  \bibinfo{author}{\bibfnamefont{J.}~\bibnamefont{Kurths}},
  \bibinfo{journal}{Phys. Rev. E} \textbf{\bibinfo{volume}{71}},
  \bibinfo{pages}{016116} (\bibinfo{year}{2005}).

\bibitem[{\citenamefont{Hwang et~al.}(2005)\citenamefont{Hwang, Chavez, Amann,
  and Boccaletti}}]{PhysRevLett.94.138701}
\bibinfo{author}{\bibfnamefont{D.-U.} \bibnamefont{Hwang}},
  \bibinfo{author}{\bibfnamefont{M.}~\bibnamefont{Chavez}},
  \bibinfo{author}{\bibfnamefont{A.}~\bibnamefont{Amann}}, \bibnamefont{and}
  \bibinfo{author}{\bibfnamefont{S.}~\bibnamefont{Boccaletti}},
  \bibinfo{journal}{Phys. Rev. Lett.} \textbf{\bibinfo{volume}{94}},
  \bibinfo{pages}{138701} (\bibinfo{year}{2005}).

\bibitem[{\citenamefont{Ipsen and Mikhailov}(2002)}]{Mikhailov02}
\bibinfo{author}{\bibfnamefont{M.}~\bibnamefont{Ipsen}} \bibnamefont{and}
  \bibinfo{author}{\bibfnamefont{A.~S.} \bibnamefont{Mikhailov}},
  \bibinfo{journal}{Phys. Rev. E} \textbf{\bibinfo{volume}{66}},
  \bibinfo{pages}{046109} (\bibinfo{year}{2002}).

\bibitem[{\citenamefont{{Moyano} et~al.}(2001)\citenamefont{{Moyano},
  {Abramson}, and {Zanette}}}]{Moyano01}
\bibinfo{author}{\bibfnamefont{L.~G.} \bibnamefont{{Moyano}}},
  \bibinfo{author}{\bibfnamefont{G.}~\bibnamefont{{Abramson}}},
  \bibnamefont{and} \bibinfo{author}{\bibfnamefont{D.~H.}
  \bibnamefont{{Zanette}}}, \bibinfo{journal}{Eur. Phys. J. B}
  \textbf{\bibinfo{volume}{22}}, \bibinfo{pages}{223} (\bibinfo{year}{2001}).

\bibitem[{\citenamefont{Yanagita and Mikhailov}(2010)}]{yanagita10}
\bibinfo{author}{\bibfnamefont{T.}~\bibnamefont{Yanagita}} \bibnamefont{and}
  \bibinfo{author}{\bibfnamefont{A.~S.} \bibnamefont{Mikhailov}},
  \bibinfo{journal}{Phys. Rev. E} \textbf{\bibinfo{volume}{81}},
  \bibinfo{pages}{056204} (\bibinfo{year}{2010}).

\bibitem[{\citenamefont{Kuramoto}(1984)}]{Kuramoto84}
\bibinfo{author}{\bibfnamefont{Y.}~\bibnamefont{Kuramoto}},
  \emph{\bibinfo{title}{Chemical Oscillations, Waves, and Turbulence}}
  (\bibinfo{publisher}{Springer}, \bibinfo{year}{1984}).

\bibitem[{\citenamefont{Acebr\'on et~al.}(2005)\citenamefont{Acebr\'on,
  Bonilla, P\'erez~Vicente, Ritort, and Spigler}}]{RevModPhys77137}
\bibinfo{author}{\bibfnamefont{J.~A.} \bibnamefont{Acebr\'on}},
  \bibinfo{author}{\bibfnamefont{L.~L.} \bibnamefont{Bonilla}},
  \bibinfo{author}{\bibfnamefont{C.~J.} \bibnamefont{P\'erez~Vicente}},
  \bibinfo{author}{\bibfnamefont{F.}~\bibnamefont{Ritort}}, \bibnamefont{and}
  \bibinfo{author}{\bibfnamefont{R.}~\bibnamefont{Spigler}},
  \bibinfo{journal}{Rev. Mod. Phys.} \textbf{\bibinfo{volume}{77}},
  \bibinfo{pages}{137} (\bibinfo{year}{2005}).

\bibitem[{\citenamefont{Mikhailov and Calenbuhr}(2006)}]{Mikhailov06}
\bibinfo{author}{\bibfnamefont{A.~S.} \bibnamefont{Mikhailov}}
  \bibnamefont{and}
  \bibinfo{author}{\bibfnamefont{V.}~\bibnamefont{Calenbuhr}},
  \emph{\bibinfo{title}{From Cells to Societies}}
  (\bibinfo{publisher}{Springer}, \bibinfo{address}{New York},
  \bibinfo{year}{2006}).

\bibitem[{\citenamefont{Landau and Binder}(2005)}]{Landau05}
\bibinfo{author}{\bibfnamefont{D.}~\bibnamefont{Landau}} \bibnamefont{and}
  \bibinfo{author}{\bibfnamefont{K.}~\bibnamefont{Binder}},
  \emph{\bibinfo{title}{A Guide to Monte Carlo Simulations in Statistical
  Physics}} (\bibinfo{publisher}{Cambridge University Press},
  \bibinfo{year}{2005}).

\bibitem[{\citenamefont{Newman and Barkema}(1999)}]{Newman99}
\bibinfo{author}{\bibfnamefont{M.~E.~J.} \bibnamefont{Newman}}
  \bibnamefont{and} \bibinfo{author}{\bibfnamefont{G.~T.}
  \bibnamefont{Barkema}}, \emph{\bibinfo{title}{Monte Carlo Methods in
  Statistical Physics}} (\bibinfo{publisher}{Oxford University Press},
  \bibinfo{year}{1999}).

\bibitem[{\citenamefont{Liu}(2001)}]{Liu01}
\bibinfo{author}{\bibfnamefont{J.}~\bibnamefont{Liu}},
  \emph{\bibinfo{title}{Monte Carlo Strategies in Scientific Computing}}
  (\bibinfo{publisher}{Springer}, \bibinfo{year}{2001}).

\bibitem[{\citenamefont{Cho et~al.}(1994)\citenamefont{Cho, Doll, and
  Freeman}}]{Cho94}
\bibinfo{author}{\bibfnamefont{A.~E.} \bibnamefont{Cho}},
  \bibinfo{author}{\bibfnamefont{J.~D.} \bibnamefont{Doll}}, \bibnamefont{and}
  \bibinfo{author}{\bibfnamefont{D.~L.} \bibnamefont{Freeman}},
  \bibinfo{journal}{Chem. Phys. Lett.} \textbf{\bibinfo{volume}{229}},
  \bibinfo{pages}{218} (\bibinfo{year}{1994}).

\bibitem[{\citenamefont{Bolhuis et~al.}(1998)\citenamefont{Bolhuis, Dellago,
  and Chandler}}]{Bolhuis98}
\bibinfo{author}{\bibfnamefont{P.~G.} \bibnamefont{Bolhuis}},
  \bibinfo{author}{\bibfnamefont{C.}~\bibnamefont{Dellago}}, \bibnamefont{and}
  \bibinfo{author}{\bibfnamefont{D.}~\bibnamefont{Chandler}},
  \bibinfo{journal}{Faraday Discuss.} \textbf{\bibinfo{volume}{110}},
  \bibinfo{pages}{421} (\bibinfo{year}{1998}).

\bibitem[{\citenamefont{Vlugt and Smit}(2000)}]{Vlugt00}
\bibinfo{author}{\bibfnamefont{T.}~\bibnamefont{Vlugt}} \bibnamefont{and}
  \bibinfo{author}{\bibfnamefont{B.}~\bibnamefont{Smit}},
  \bibinfo{journal}{ChemComm.} \textbf{\bibinfo{volume}{2}},
  \bibinfo{pages}{11} (\bibinfo{year}{2000}).

\bibitem[{\citenamefont{Kawasaki and Sasa}(2005)}]{Kawasaki05}
\bibinfo{author}{\bibfnamefont{M.}~\bibnamefont{Kawasaki}} \bibnamefont{and}
  \bibinfo{author}{\bibfnamefont{S.~I.} \bibnamefont{Sasa}},
  \bibinfo{journal}{Phys. Rev. E} \textbf{\bibinfo{volume}{72}},
  \bibinfo{pages}{037202} (\bibinfo{year}{2005}).

\bibitem[{\citenamefont{Sasa and Hayashi}(2006)}]{Sasa06}
\bibinfo{author}{\bibfnamefont{S.~I.} \bibnamefont{Sasa}} \bibnamefont{and}
  \bibinfo{author}{\bibfnamefont{K.}~\bibnamefont{Hayashi}},
  \bibinfo{journal}{Europhys. Lett.} \textbf{\bibinfo{volume}{74}},
  \bibinfo{pages}{156} (\bibinfo{year}{2006}).

\bibitem[{\citenamefont{Giardin\'{a} et~al.}(2006)\citenamefont{Giardin\'{a},
  Kurchan, and Peliti}}]{Giardin06}
\bibinfo{author}{\bibfnamefont{C.}~\bibnamefont{Giardin\'{a}}},
  \bibinfo{author}{\bibfnamefont{J.}~\bibnamefont{Kurchan}}, \bibnamefont{and}
  \bibinfo{author}{\bibfnamefont{L.}~\bibnamefont{Peliti}},
  \bibinfo{journal}{Phys. Rev. Lett.} \textbf{\bibinfo{volume}{96}},
  \bibinfo{pages}{120603} (\bibinfo{year}{2006}).

\bibitem[{\citenamefont{Tailleur and Kurchan}(2007)}]{Tailleur07}
\bibinfo{author}{\bibfnamefont{J.}~\bibnamefont{Tailleur}} \bibnamefont{and}
  \bibinfo{author}{\bibfnamefont{J.}~\bibnamefont{Kurchan}},
  \bibinfo{journal}{Nature Physics} \textbf{\bibinfo{volume}{3}},
  \bibinfo{pages}{203} (\bibinfo{year}{2007}).

\bibitem[{\citenamefont{Yanagita and Iba}(2009)}]{yanagita09}
\bibinfo{author}{\bibfnamefont{T.}~\bibnamefont{Yanagita}} \bibnamefont{and}
  \bibinfo{author}{\bibfnamefont{Y.}~\bibnamefont{Iba}}, \bibinfo{journal}{J.
  Stat. Mech.} \textbf{\bibinfo{volume}{2}}, \bibinfo{pages}{02043}
  (\bibinfo{year}{2009}).

\bibitem[{\citenamefont{Teramae and Tanaka}(2004)}]{Teramae04}
\bibinfo{author}{\bibfnamefont{J.-n.} \bibnamefont{Teramae}} \bibnamefont{and}
  \bibinfo{author}{\bibfnamefont{D.}~\bibnamefont{Tanaka}},
  \bibinfo{journal}{Phys. Rev. Lett.} \textbf{\bibinfo{volume}{93}},
  \bibinfo{pages}{204103} (\bibinfo{year}{2004}).

\bibitem[{\citenamefont{Nakao et~al.}(2007)\citenamefont{Nakao, Arai, and
  Kawamura}}]{Nakao07}
\bibinfo{author}{\bibfnamefont{H.}~\bibnamefont{Nakao}},
  \bibinfo{author}{\bibfnamefont{K.}~\bibnamefont{Arai}}, \bibnamefont{and}
  \bibinfo{author}{\bibfnamefont{Y.}~\bibnamefont{Kawamura}},
  \bibinfo{journal}{Phys. Rev. Lett.} \textbf{\bibinfo{volume}{98}},
  \bibinfo{pages}{184101} (\bibinfo{year}{2007}).

\bibitem[{\citenamefont{Kori et~al.}(2012)\citenamefont{Kori, Kawamura, and
  Masuda}}]{Kori201261}
\bibinfo{author}{\bibfnamefont{H.}~\bibnamefont{Kori}},
  \bibinfo{author}{\bibfnamefont{Y.}~\bibnamefont{Kawamura}}, \bibnamefont{and}
  \bibinfo{author}{\bibfnamefont{N.}~\bibnamefont{Masuda}},
  \bibinfo{journal}{Journal of Theoretical Biology}
  \textbf{\bibinfo{volume}{297}}, \bibinfo{pages}{61 } (\bibinfo{year}{2012}).

\bibitem[{\citenamefont{Masuda et~al.}(2010)\citenamefont{Masuda, Kawamura, and
  Kori}}]{Masuda10}
\bibinfo{author}{\bibfnamefont{N.}~\bibnamefont{Masuda}},
  \bibinfo{author}{\bibfnamefont{Y.}~\bibnamefont{Kawamura}}, \bibnamefont{and}
  \bibinfo{author}{\bibfnamefont{H.}~\bibnamefont{Kori}}, \bibinfo{journal}{New
  Journal of Physics} \textbf{\bibinfo{volume}{12}}, \bibinfo{pages}{093007}
  (\bibinfo{year}{2010}).

\end{thebibliography}
\end{document}